\documentclass[aps,pra,twocolumn,groupedaddress]{revtex4-1}

\usepackage{graphicx}
\usepackage{amsmath}
\usepackage{latexsym}		
\usepackage{graphics}
\usepackage[usenames]{color}
\usepackage{color}
\usepackage{array}
\usepackage{enumerate}

\begin{document}

\newcommand{\diff}[2]{\frac{\partial #1}{\partial #2}}
\newcommand{\secdiff}[2]{\frac{\partial^2 #1}{\partial #2^2}}
\newcommand{\comment}[1]{{\bf \textcolor{red}{ #1}}}
\newcommand{\BDEcomment}[1]{{\bf \textcolor{blue}{ #1}}}
\newcommand{\NPMcomment}[1]{{\bf \textcolor{ForestGreen}{ #1}}}

\newcounter{tabitem}

\title{Few-Boson Processes in the Presence of an Attractive Impurity under One-Dimensional Confinement}

\author{Nirav P. Mehta}
\email[]{nmehta@trinity.edu}
\affiliation{Trinity University, San Antonio, TX. 78212-7200 USA}
\author{Connor D. Morehead}
\affiliation{Trinity University, San Antonio, TX. 78212-7200 USA}

\date{\today}

\begin{abstract}
We consider a few-boson system confined to one dimension
with a single distinguishable particle of lesser mass. 
All particle interactions are modeled with
$\delta$-functions, but due to the mass imbalance the problem is
nonintegrable. Universal few-body binding energies, atom-dimer and
atom-trimer scattering lengths are all calculated in terms of two parameters,
namely the mass ratio: $m_{\text{L}}/m_{\text{H}}$, and ratio
$g_{\text{HH}}/g_{\text{HL}}$ of the $\delta$-function 
couplings. We specifically identify the values of these ratios for
which the atom-dimer or atom-trimer scattering lengths vanish or
diverge. We identify regions in this parameter space in which various
few-body inelastic process become energetically allowed. 
In the Tonks-Girardeau limit ($g_{\text{HH}}\rightarrow \infty$), our results are relevant to experiments
involving trapped fermions with an impurity atom.
\end{abstract}


\maketitle

\section{Introduction}

Strongly interacting one dimensional (1D) quantum systems have been
of fundamental interest for many
years~\cite{Cazalilla2011RMP,Guan2013RMP}.  Not long ago, a number of
experiments~\cite{Paredes2004Nature,Moritz2003PRL,Kinoshita2006Nature}
involving quantum gases tightly confined to two-dimensional (2D)
optical lattices realized the 1D Lieb-Liniger-McGuire
model~\cite{LiebLiniger1963PhysRev,Lieb1963PhysRev,McGuire1964JMP}.
Interpretation of these experiments in terms of the one-dimensional
model parameters has been facilitated by a
calculation of the effective 1D
coupling constant ($g_{\text{1D}}$) and dimer energy ($E_2$) in
terms of the three-dimensional (3D) s-wave scattering length
($a_{\text{3D}}$) and the transverse confinement length ($a_\perp=(\mu\omega_\perp)^{-1/2}$)~\cite{Olshanii1998PRL,Bergeman2003PRL} (in units with
$\hbar = 1$ throughout): 
\begin{equation}
\label{Olshanii}
g_{\text{1D}}=\frac{2 a_{\text{3D}}}{\mu a_\perp^2 (1-\left|\zeta\left(\frac{1}{2}\right)\right|
\frac{a_{\text{3D}}}{a_\perp})}\text{;}\;\;
\zeta\left(\frac{1}{2}, \Omega_B\right)=-\frac{a_\perp}{a_{\text{3D}}}
\end{equation}
where $\zeta(z,q)$ is the generalized Riemann zeta
function~\cite{GradshteynRyzhik}, $\mu$ is the
two-body reduced mass, and
$\Omega_B=1/2-E_2/(2\omega_\perp)$ is the dimensionless
dimer binding energy.  The result predicts the
so-called ``confinement induced resonance'' (CIR) when $|\zeta(1/2)|
a_{\text{3D}} = a_\perp$, where $|g_{\text{1D}}|\rightarrow \infty$
and the 1D scattering length $a_{\text{1D}} = -(\mu
g_{\text{1D}})^{-1}$ passes through zero.  Remarkably, $g_{\text{1D}}$
can be experimentally tuned by varying either $a_{\text{3D}}$ (via a magnetic Feshbach
resonance), or $a_\perp$.  

Moreover, equation~(\ref{Olshanii}) predicts that a 1D dimer always exists
below the asymptotic threshold energy $\omega_\perp$ regardless
of the sign of $a_{\text{3D}}$.  Such confinement-induced dimers
have been seen in Monte Carlo simulations~\cite{Astrakharchik2004PRL,Astrakharchik2004JPhysB} and
observed experimentally by RF spectroscopy~\cite{Moritz2005PRL}.  When
the axial extent of the dimer is large compared to the transverse
confinement length, the few-body physics is expected to follow from a
purely 1D calculation with $\delta$-function couplings given by
Eq.~(\ref{Olshanii})~\cite{Astrakharchik2004PRL,Astrakharchik2004JPhysB,Mora2005PRA,Mora2005PRL}.
This limit is achieved when $a_\perp/a_{\text{3D}}\rightarrow
-\infty$, such that $\Omega_B \ll 1$. 
 
More recent quasi-1D experiments have begun to probe the interaction of a
degenerate quantum gas with a distinguishable
impurity~\cite{Moritz2005PRL,Wenz2013Science,Catani2012PRA,Pagano2014NatPhys}.
Motivated by the aforementioned evidence that---in the appropriate limit---purely 1D calculations of few-body
processes can provide physically meaningful insights into quasi-1D
experiments, we consider here the $N=3$ and $N=4$ instances of the following Hamiltonian:
\begin{equation}
\label{H1}
  \begin{split}
    H &= -\frac{1}{2m_{\text{H}}} \sum_{i=1}^{N-1}{\secdiff{}{x_i}}
    -\frac{1}{2m_{\text{L}}}\secdiff{}{x_N} \\
    &+\sum_{i<j}^{N-1}{g_{\text{HH}}\delta(x_i-x_j)} +
    \sum_{i=1}^{N-1}{g_{\text{HL}}\delta(x_i - x_N)}, 
  \end{split}
\end{equation}
where particles $1$ through $(N-1)$ are identical bosons (H) of mass
$m_{\text{H}}$, and particle $N$ is an impurity (L) of mass $m_{\text{L}}$. 
The equal-mass ($m_{\text{L}}=m_{\text{H}}$) instance of Eq.~(\ref{H1})
has been realized with $^{40}\text{K}$ atoms~\cite{Moritz2005PRL} and
more recently $^6\text{Li}$~\cite{Wenz2013Science}, with related
theory work found in Refs.~\cite{Gharashi2013PRL,Gharashi2012PRA,Sowinski2013PRA,Gharashi2015PRA,Lindgren2014NJP,Volosniev2014NatureComm,Levinsen2015ScienceAdvances,Dehkharghani2015Preprint}.  Note that the
Bose-Fermi mapping~\cite{Girardeau1960JMP} allows one to consider the
$g_{\text{HH}}\rightarrow \infty$ instance of Eq.~(\ref{H1}) for a
description of a noninteracting background gas of fermions.  The unequal mass case
($m_{\text{H}} > m_{\text{L}}$) of Eq.~(\ref{H1}) has been realized with a
$^{41}\text{K}$ impurity in a background gas of 
$^{87}\text{Rb}$~\cite{Catani2012PRA}.  

We assume that the impurity is of lesser mass than the background
atoms, and employ the Born-Oppenheimer method to obtain adiabatic
potentials for the heavy-particle motion.  In the thermodynamic limit, stability of the background gas requires
$g_{\text{HH}}>0$.  At the few-body level, our calculation permits us to consider
$g_{\text{HH}}<0$, but with the caveat that the lowest
scattering threshold involves a bound impurity. 
The eigenstates of Eq.~(\ref{H1}) are completely specified by the coupling
ratio $\lambda = g_{\text{HH}}/g_{\text{HL}}$ and the mass ratio
$\beta = m_{\text{L}}/m_{\text{H}}$. The case $N=3$ of Eq.~(\ref{H1})
has been studied in considerable detail by Kartavtsev \emph{et
  al}~\cite{Kartavtsev2009JETP}, who treat all possible values of
$\lambda$ and $\beta$ using the adiabatic hyperspherical
representation.  Mehta~\cite{Mehta2014PRA} used the Born-Oppenheimer
method to extend those results to the case $N=4$, but only considered the cases
$\lambda \rightarrow 0$ and $|\lambda| \rightarrow \infty$ with
$\beta\le 1$. This work is an extension of those calculations to
finite values of $\lambda$. Some of the approximate $N=3$ calculations presented here can be
directly compared to the benchmark calculations
of~\cite{Kartavtsev2009JETP}, but we are unaware of
similar calculations for $N=4$.  

We calculate the atom-dimer scattering length $a_{\text{AD}}$, the
atom-trimer scattering length $a_{\text{AT}}$ and trimer and tetramer
bound-state energies as a function of $\beta$ and $\lambda$.  The
critical values of $\beta$ and $\lambda$ at which these scattering
lengths diverge are identified and marked by the appearance of a new
bound state.  We find that few-body bound states tend towards deeper
binding as one increases $\lambda$ (making it less
negative) or increases $\beta^{-1}$.  This behavior allows us to 
identify the critical values of $\lambda$ and $\beta$ where particular
inelastic processes become energetically allowed.

\section{Born-Oppenheimer solution}
Here, we briefly sketch the Born-Oppenheimer calculation.  Many elements of the
derivation are unchanged from~\cite{Mehta2014PRA}, so we present an
abridged derivation sufficient to  
highlight changes due to choosing finite
$\lambda=g_{\text{HH}}/g_{\text{HL}}$.

We scale the Hamiltonian by the HL binding energy, $B_2 =
\mu_{\text{HL}} g_{\text{HL}}^2/2 = 1/(2\mu_{\text{HL}} a^2)$. and use the HL scattering length
$a = -1/(\mu_{\text{HL}}g_{\text{HL}}) > 0$ as the fundamental length
unit.  The calculation of few-body observables in a homogeneous system begins
the removal of the center-of-mass motion by a transformation to
mass-scaled Jacobi coordinates: $\{x,y,z\}$:.  
\begin{equation}
\begin{split}
\label{JacobiCoors}
x &= \frac{1}{a}\sqrt{\frac{\mu_{12}}{\mu_{4b}}}\left(x_1 - x_2 \right) \\
y &= \frac{1}{a}\sqrt{\frac{\mu_{12,3}}{\mu_{4b}}}\left(\frac{m_1 x_1 + m_2 x_2}{m_1+m_2}-x_3 \right) \\
z &= \frac{1}{a}\sqrt{\frac{\mu_{123,4}}{\mu_{4b}}}\left(\frac{m_1 x_1
    + m_2 x_2 + m_3 x_3}{m_1 + m_2 + m_3}-x_4 \right).
\end{split}
\end{equation}
Here, $\mu_{4b}=(\mu_{12}\mu_{12,3}\mu_{123,4})^{1/3}$ is the
four-body reduced mass, and $\mu_{a,b}$ is a two-body reduced mass
between clusters $a$ and $b$.  The coordinates for the three-particle
problem are the same, but with the $z$-coordinate omitted and $\mu_{4b}$
replaced by $\mu_{3b}=(\mu_{12}\mu_{12,3})^{1/2}$.  
\subsection{Three-Body Problem}
The Born-Oppenheimer factorization of the three-body wavefunction
$\Psi(x,y)=\Phi(x;y)\psi(x)$ requires that $\Phi(x;y)$ be an
eigenstate of the fixed-$x$ Hamiltonian:
\begin{equation}
\label{Had3bod}
H_{ad}^{(3)}=-\frac{1}{2\mu_3}\secdiff{}{x}+g_3\left[\delta(y-x_0) + \delta(y+x_0)\right],
\end{equation}
with $x$-dependent eigenvalue $u(x)$.  We have defined the unitless three-body reduced mass $\mu_3 =
(1+\beta)/[2\sqrt{\beta(2+\beta)}]$, coupling $g_3 = -2\sqrt{2}
[\beta/(2+\beta)]^{1/4}$, and
scaled heavy-particle separation $x_0 = x
\sqrt{\beta/(2+\beta)}$. The solution to Eq.~(\ref{Had3bod}) yields the following
transcendental equation for the lowest $u(x)$:
\begin{equation}
\label{HHLtraneq}
\frac{\kappa}{g_3 \mu_3} + 1 = - \exp(-2\kappa x_0),
\end{equation}
where $\kappa = \sqrt{-2\mu_{\text{HL}}u(x)}$. The heavy-particle wavefunction $\psi(x)$ is now governed by the effective Hamiltonian:
\begin{equation}
\label{HHEffEq}
H_{\text{eff}}^{(3)}=\frac{-1}{2\mu_3}\secdiff{}{x}+g_3\lambda\delta(2x_0)+u(x)+\frac{\tilde{Q}(x)}{2\mu_3},
\end{equation}
Here, $\tilde{Q}(x) = \left \langle\Phi^\prime \right. \left |
  \Phi^\prime \right \rangle_y$ is the diagonal nonadiabatic 
correction, where the primes denote derivatives with respect to the
slow coordinate $x$, and the
integration is carried out over the fast coordinate $y$ only. 
Ignoring $\tilde{Q}(x)$ leads to a lower bound on the three-body
ground-state energy, while including $\tilde{Q}(x)$ leads to an upper bound that is
typically more accurate~\cite{Brattsev1965Dokl,Epstein1966JCP,Strace1979PRA,Coelho1991PRA}.  We refer to the former calculation as the
``extreme adiabatic approximation'' (EAA) and the latter as the
``uncoupled adiabatic approximation'' (UAA).
The cases $\lambda=0$ and $|\lambda|\rightarrow \infty$ were treated
in~\cite{Mehta2014PRA}.  Here we consider arbitrary $\lambda$, and
replace the $\delta$-function in Eq.~(\ref{HHEffEq}) with the following boundary condition on the
solution $\psi(x)$ for $x>0$:
\begin{equation}
\label{NewpsiBC}
\left(\frac{1}{\psi}\diff{\psi}{x}\right)_{x\rightarrow 0^+} = -\lambda\frac{1+\beta}{\sqrt{2\beta}[\beta(2+\beta)]^{1/4}}
\end{equation}

\subsection{Four-Body Problem}
For the four-particle problem, it is convenient to work in ``hyper-cylindrical''
coordinates, trading $\{x,y,z\} \rightarrow \{\rho,\phi,z\}$ by the
usual transformation: $\rho^2 = x^2+y^2$, and $\tan{\phi}=y/x$. We
make the Born-Oppenheimer factorization: $\Psi(\rho,\phi,z) =
\Phi(\rho,\phi;z)\psi(\rho,\phi)$, and integrate out the
light-particle degree of freedom by demanding $\Phi(\rho,\phi;z)$ be
an eigenstate of the following fixed-$\{\rho,\phi\}$ Hamiltonian with eigenvalue $U(\rho,\phi)$:
\begin{equation}
\label{HHHLAdSE}
H_{ad}^{(4)}=\frac{-1}{2\mu_4}\secdiff{}{z}+g_4\sum_{i=1}^{3}{\delta(z-z_i)}.
\end{equation}
We have again introduced unitless parameters: $\mu_4 = (\beta +1)/[2 \beta ^{2/3}
  (3+\beta)^{1/3}]$, and $g_4 = -2 \sqrt{3}
  \left[\beta/(3+\beta)\right]^{1/3}$. The heavy-light coalescence points occur at
$z_i=\sqrt{2\beta/({3+\beta})} \; \rho \sin \left(\phi -
  \phi_i\right)$ with $\phi_1 = -4\pi/3$, $\phi_2 = 0$, and $\phi_3 = -2\pi/3$.
The adiabatic equation is simply a triple-$\delta$ problem that leads
to the following transcendental equation for the potential energy surface:
\begin{equation}
\label{HHHLtraneq}
\frac{(g+2 \kappa )^2 }{g^2}e^{2 \kappa  z_3}+\frac{(g-2 \kappa ) }{g+2 \kappa }e^{2 \kappa  z_1}=e^{2 \kappa  z_2}+e^{2 \kappa  \left(z_1-z_2+z_3\right)}
\end{equation}
where $g=2\mu_4g_4$, $\kappa^2 = -2\mu_4 U(\rho,\phi)>0$.

The heavy-particle eigenstates $\psi(\rho,\phi)$ now live on the potential energy
surface $U(\rho,\phi)$, and are eigenstates of the effective Hamiltonian (in the EAA):
\begin{equation}
\label{BOEquation}
\begin{split}
H_{\text{eff}}^{(4)}=&\frac{-1}{2\mu_4}\left(\frac{1}{\rho}\diff{}{\rho}\rho\diff{}{\rho}+\frac{1}{\rho^2}\secdiff{}{\phi}\right)
\\
&+\left[U(\rho,\phi)
  + \frac{\lambda g_4}{\alpha \rho} \sum_{i<j}^{3}{\delta\left(\left|\sin(\phi-\phi_{ij})\right|\right)}\right]
\end{split}
\end{equation}
where $\alpha = \sqrt{6\beta/(3+\beta)}$.  In order to extract scattering lengths
we require a representation of the wavefunction that gives the
appropriate asymptotic cluster states.  We write: $\psi(\rho,\phi) =
\sum_{n=0}^{\infty}{\rho^{-1/2}f_n(\rho)\chi_n(\rho;\phi)}$, where the
channel functions $\chi_n$ satisfy the fixed-$\rho$
equation:
\begin{equation}
\label{fixedrhoeq}
\frac{-1}{2\mu_4\rho^2}\secdiff{\chi_n}{\phi}+U(\rho,\phi)\chi_n =  U_n(\rho)\chi_n
\end{equation}
Due to identical particle symmetry, one only needs to consider the
restricted range $\phi \in [0,\pi/6)$ with the boundary condition~\cite{Mehta2014PRA}:
\begin{equation}
\label{NewChiBC}
\lim_{\epsilon \rightarrow
  0}{\left.\frac{1}{\chi}\diff{\chi}{\phi}\right|_{\pi/6-\epsilon}}=\frac{\rho
    \lambda (1+\beta)}{\sqrt{2}\beta}\left(\frac{\beta}{3+\beta}\right)^{1/6}
\end{equation}
Finally, the four-particle problem is
reduced to a set of coupled equations in $\rho$ only:
\begin{equation}
\label{hyperradialeqs}
\begin{split}
\frac{-1}{2\mu_4}&\left(\mathbf{1}\secdiff{}{\rho}+\mathbf{Q}(\rho)+2\mathbf{P}(\rho)\diff{}{\rho}\right)\vec{f}(\rho)\\
&+\mathbf{U}_{\text{eff}}(\rho)\vec{f}(\rho)=E_{EAA}\vec{f}(\rho)
\end{split}
\end{equation}
Here, $\mathbf{U}_{\text{eff}}$ is a diagonal matrix with elements $U_n(\rho)
- 1/8\mu_4\rho^2$, $P_{mn}(\rho)=\left\langle
  \chi_m\left|\chi_n^\prime\right.\right\rangle_\phi$ and $Q_{mn}(\rho)=\left\langle \chi_m
  \left|\chi_n^{\prime\prime}\right.\right\rangle_\phi$.  The Born-Oppenheimer potentials obtained from Eqs.~(\ref{HHLtraneq})
and~(\ref{HHHLtraneq}) are independent of $\lambda$, and only need to
be calculated once for a given $\beta$.  The $\lambda$-dependence
only appears through the boundary conditions~(\ref{NewpsiBC})
and~(\ref{NewChiBC}).  

In Table~\ref{table:BCs}, we sumarize the boundary conditions on the
channel functions $\chi_n(\rho;\phi)$.  Note that starting with
positive parity noninteracting bosons, one achieves the
infinitely repulsive ``fermionized'' limit as $\lambda \rightarrow -\infty$ (since
$g_{\text{HL}}<0$).  Equation~(\ref{NewChiBC}) then gives a boundary
condition identical to that of noninteracting fermions, but of
negative parity.  Due to Pauli exclusion, identical (spin-polarized) fermions are
insensitive to the zero-range interaction, and the boundary
conditions for fermions are unaffected by $\lambda$.

\begin{table}[!h]
  \caption{
    \label{table:BCs}
    The boundary conditions on the channel functions $\chi_n(\rho;\phi)$
    for the non-interacting (NI), infinitely repulsive (IR) cases are
    summarized here.  
  }
  \begin{center}
    \begin{tabular}{|c|c|c|}
      \hline
      & $(+)$ parity & $(-)$ parity \\
      \hline
      \hline
      NI bosons &
      \shortstack{$\left[\diff{\chi}{\phi}\right]_{\phi=0}=0$\\
      $\left[\diff{\chi}{\phi}\right]_{\phi=\pi/6}=0$} & 
    \shortstack{$\left[\diff{\chi}{\phi}\right]_{\phi=0}=0$
    \\$\left.\chi\right|_{\phi=\pi/6}=0$} \\
  \hline
  IR bosons &       \shortstack{$\left[\diff{\chi}{\phi}\right]_{\phi=0}=0$\\
    $\left.\chi\right|_{\phi=\pi/6}=0$} & 
  \shortstack{$\left.\chi\right|_{\phi=0}=0$
    \\$\left.\chi\right|_{\phi=\pi/6}=0$} \\
  \hline
  NI fermions &  \shortstack{$\left.\chi\right|_{\phi=0}=0$
    \\$\left.\chi\right|_{\phi=\pi/6}=0$} &
  \shortstack{$\left[\diff{\chi}{\phi}\right]_{\phi=0}=0$\\
    $\left.\chi\right|_{\phi=\pi/6}=0$} \\ 
  \hline
    \end{tabular}
  \end{center}
\end{table}

\section{Numerical Solutions}
\begin{figure}[]
\begin{center}
\leavevmode
  \includegraphics[width=3.0in,clip=true]{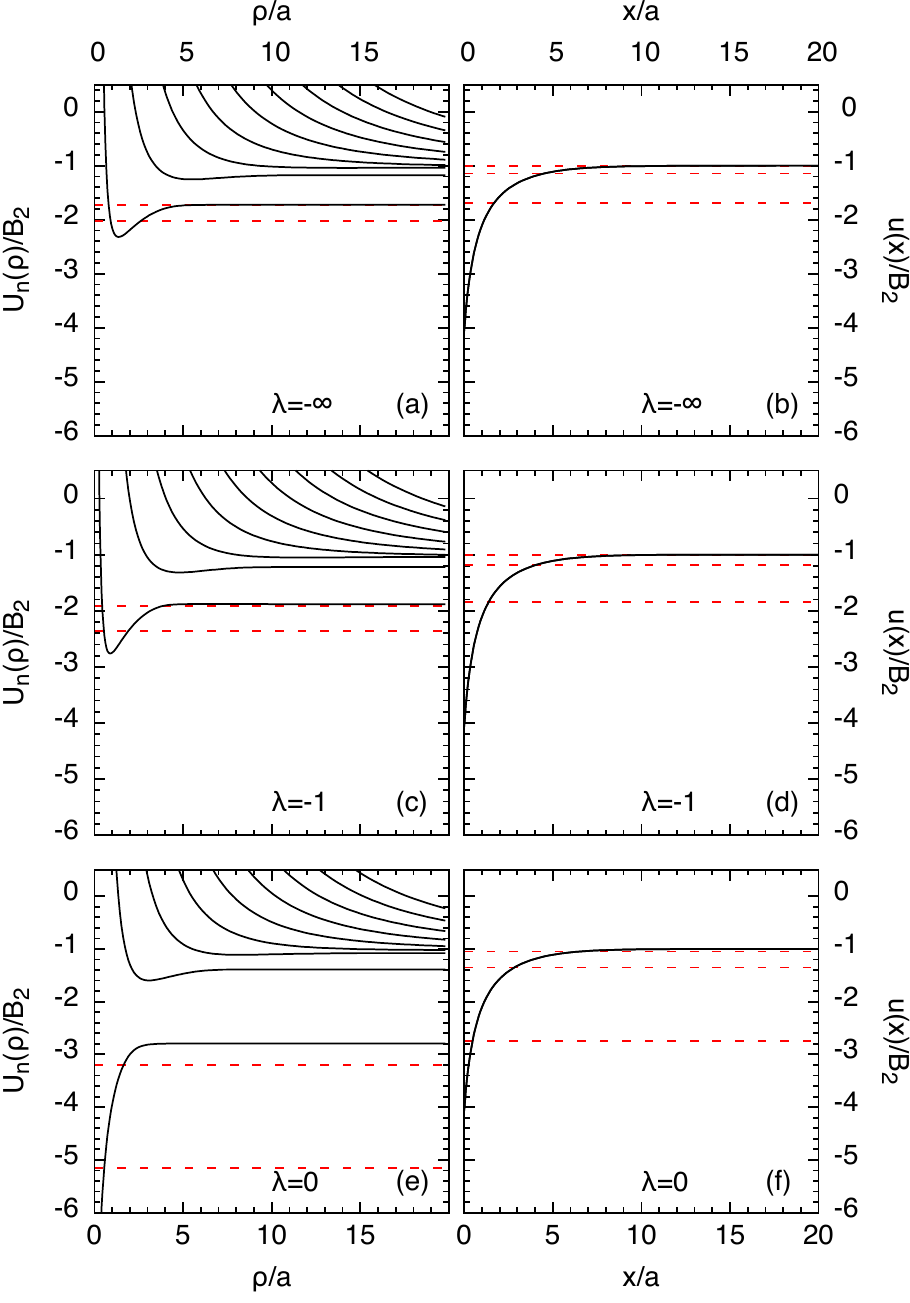}
\caption{(color online)
Panels (a), (c), and (d) show hyperradial potential energy curves
$U_n(\rho)$ determined by solving Eq.~(\ref{fixedrhoeq}) for
$\lambda\rightarrow -\infty$, $\lambda=-1$ and $\lambda=0$, respectively.  Panels (b),
(c) and (d) show the corresponding potentials $u(x)$ of Eq.~(\ref{Had3bod}).  The
dashed red lines indicate the energies of trimer and tetramer bound states.
}
\label{potCurves}
\end{center}
\end{figure}

Born-Oppenheimer and hyperradial potential curves
provide a view of the few-body energy landscape that 
aids in the interpretation subsequent calculations.  In
Fig.~\ref{potCurves}(a-b), we show potential curves and bound state
energies for the Li-Cs mass ratio $\beta^{-1/2}=4.7$, and $\lambda
\rightarrow -\infty$.  Panel (a) shows the hyperradial potential
curves $U_n(\rho)$ while panel (b) shows the effective H-H interaction
$u(x)$.  Trimer (HHL) and tetramer (HHHL) bound states are indicated
by dashed red lines in panel (a) and (b), respectively.  Panels (b-c)
and (d-e) are identical to (a-b), but for
$\lambda=-1$ and $\lambda=0$, respectively.  
Note that the potential curves $U_n(\rho)$ depend on $\lambda$ through
the boundary condition~(\ref{NewChiBC}). While $u(x)$ is independent
of $\lambda$, the trimer binding energies depend on $\lambda$ through Eq.~(\ref{NewpsiBC}).

Note that in the limit $\rho\rightarrow \infty$, the lowest potentials
$U_n(\rho)$ approach the trimer energies, asymptotically
representing atom-trimer channels (H+HHL).  In the lowest atom-trimer
channel we clearly see the appearance of a short-range potential energy barrier as
$\lambda \rightarrow -\infty$.  This barrier is a direct result of the
boundary condition~(\ref{NewChiBC}), and represents the repulsive
effect of fermionization.  
We also see a set of
potentials asymptotically approaching the dimer energy $U_n/B_2 = -1$.
These represent the three-body (H+H+HL) channels.  Three-body
recombination at threshold is controlled by the lowest such potential, which in the large
$\rho$ limit behaves as $U_n(\rho)\rightarrow
\kappa_\text{min}^2/(2\mu_4\rho^2)$.  A full hyperspherical
calcualtion would also produce a set of potentials
approaching the zero-energy threshold asymptotically representing four
free atoms (H+H+H+L).  Our calculation cannot capture these potentials
because the light particle is required to be bound.

Numerical solutions to
Eqs.~(\ref{HHEffEq}), (\ref{fixedrhoeq}) and
(\ref{hyperradialeqs}) are found by expressing the desired wave
functions as a sum over b-splines and solving the resulting
generalized eigenvalue problem.  We extract $a_{\text{AD}}$ and $a_{\text{AT}}$ by matching the solutions
$\psi(x)$ and $f_0(\rho)$ to a phase-shifted cosine in the asymptotic
region, and extrapolating the zero-energy limit of $k \tan\delta
\rightarrow 1/a$ (see~\cite{Mehta2014PRA} for details.) 

\subsection{Scattering lengths $a_{\text{AD}}$ and $a_\text{AT}$}
\begin{figure}[]
\begin{center}
\leavevmode
  \includegraphics[width=3.0in,clip=true]{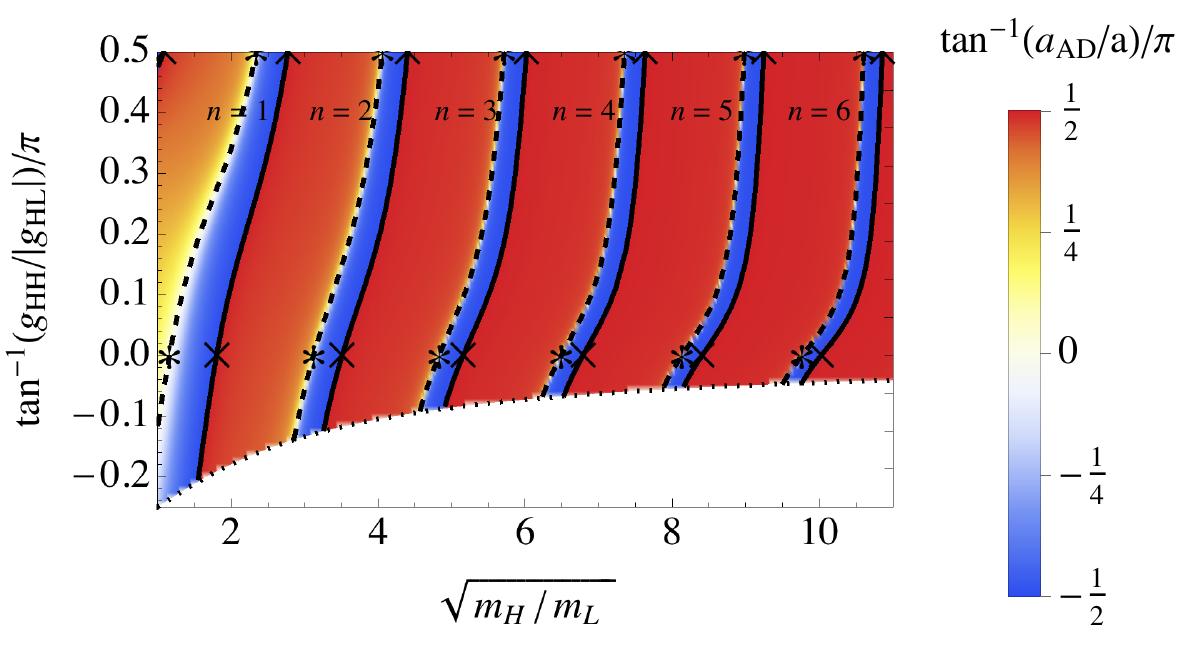}\\
  \includegraphics[width=3.0in,clip=true]{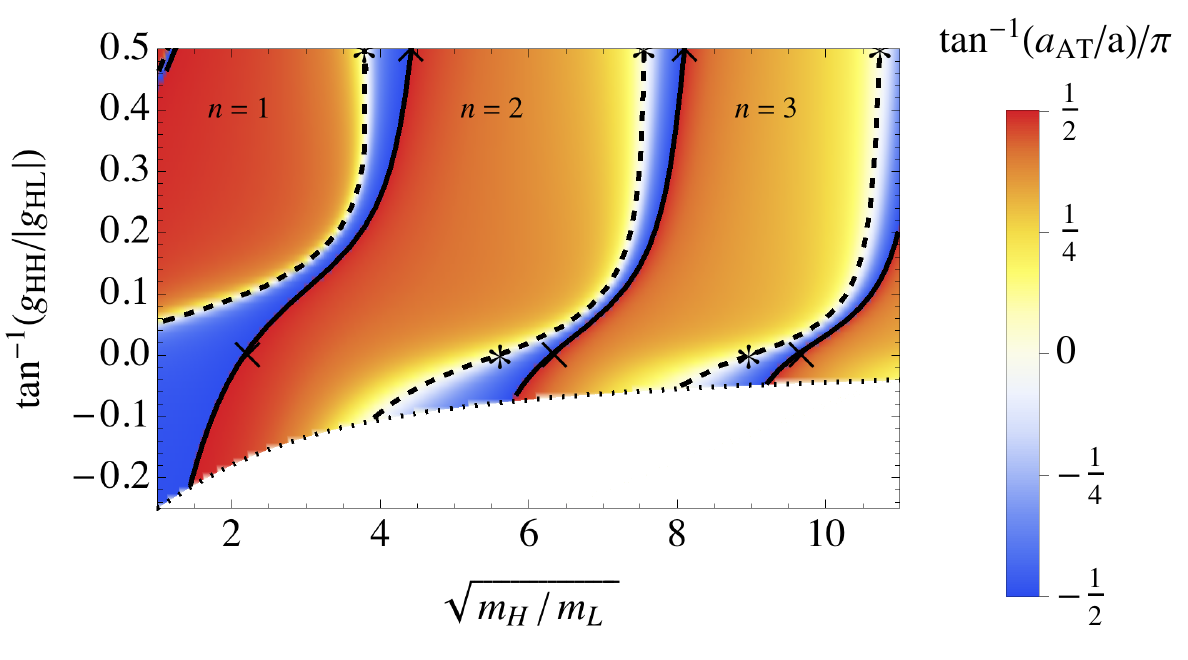}
 \begin{picture}(0,0)(0,0)
      \put(-220,235){(a)}
      \put(-220,110){(b)}
 \end{picture}
\caption{(color online)
      Panels (a) and (b) show contour plots of
      $a_{\text{AD}}$ and $a_{\text{AT}}$, respectively, on
      the plane formed by $\tan^{-1}{(g_{\text{HH}}/|g_{\text{HL}}|)}$ and
      $\sqrt{m_{\text{H}}/m_{\text{L}}}$.  The dashed lines mark the
      locus of points where the corresponding scattering length
      vanishes, while the solid lines indicate where it diverges.  Along the dotted line,
      the (HH) binding energy is equal to the (HL) binding.
}
\label{HHLphase}
\end{center}
\end{figure}
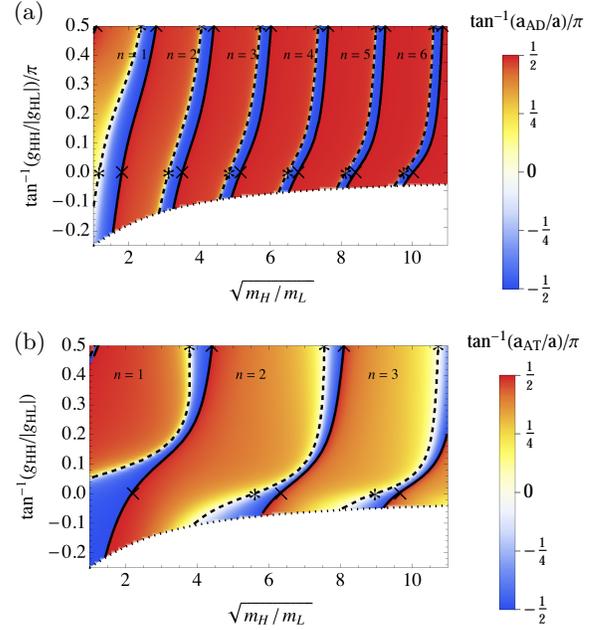

In Fig.~\ref{HHLphase}, we show contour plots of the atom-dimer
scattering length $a_{\text{AD}}$ and the 
atom-trimer scattering length $a_{\text{AT}}$ on the plane formed by $\tan^{-1}{(-\lambda)}$ and
$\beta^{-1/2}$.  We consider only the lowest atom-trimer channel for
the elastic collision $\text{H}+\text{HHL}\rightarrow 
\text{H}+\text{HHL}$ when calculating $a_{\text{AT}}$.  All higher
scattering channels are energetically closed for all $\rho$, and their
effect is negligible in comparison to the EAA.  The Born-Oppenheimer method 
requires the light particle to be bound 
such that the lowest atom-dimer scattering channel asymptotically contains
an HL dimer, describing the elastic collision: $\text{H+HL}\rightarrow
\text{H+HL}$.  Our calculation fails below the dotted line in
Figs.~\ref{HHLphase}(a) and (b), which is described by the formula $\lambda =
\sqrt{2\beta/(1+\beta)}$.  Below this line, the lowest atom-dimer
scattering threshold contains an HH dimer and a free L atom, and the
structure of the three-body phase diagram is fundamentally
different~\cite{Kartavtsev2009JETP}.  

\begin{figure}[ht]
\begin{center}
\leavevmode
\includegraphics[width=3.0in,clip=true]{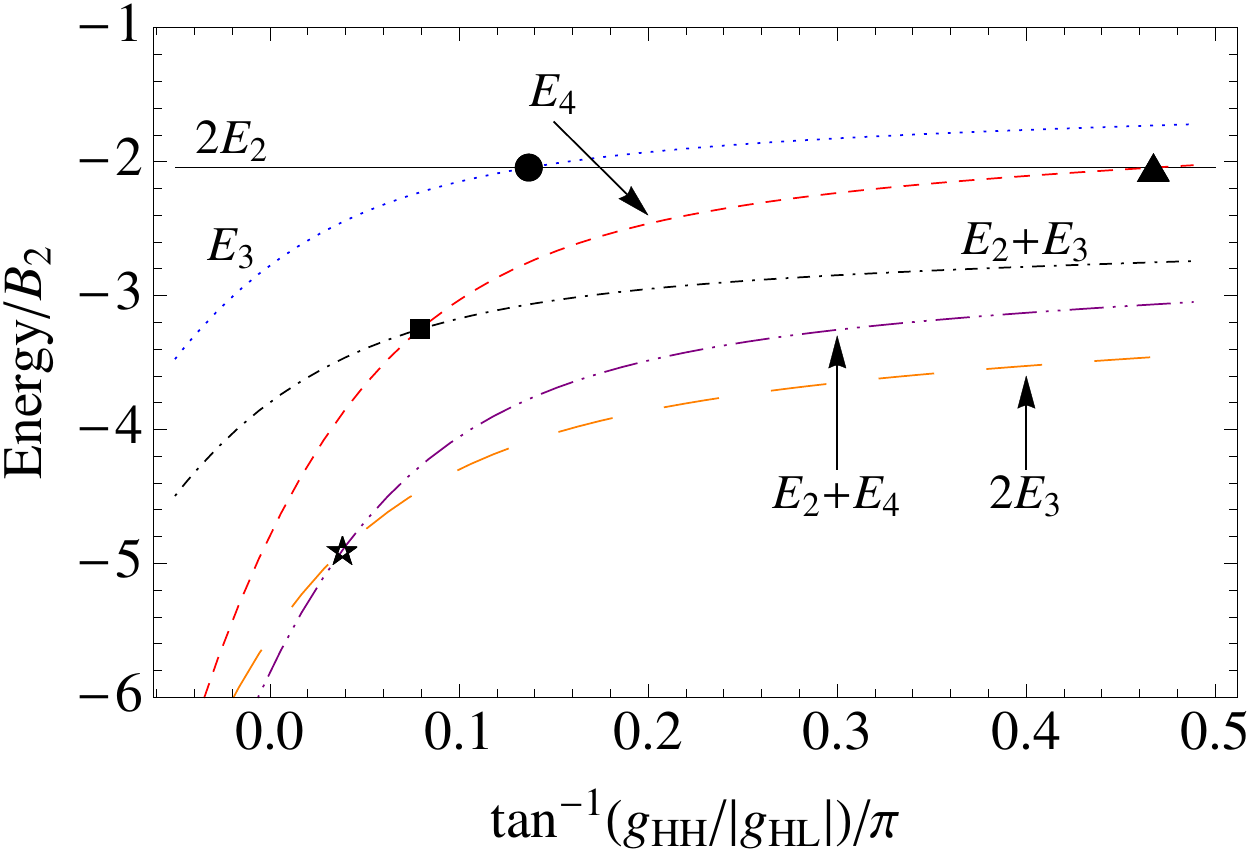}
\caption{(color online) The $\lambda$-dependence of various energies involving HL,
  HHL, and HHHL clusters for $\beta^{-1/2} = 4.7$, corresponding to
  the Li-Cs mass ratio.  All energies shown are calculated in the EAA.}
\label{pThresholds}
\end{center}
\end{figure}

The dashed lines in Figs.~\ref{HHLphase}(a) and (b) indicate
$a_{\text{AD}}\rightarrow 0$ and $a_{\text{AT}}\rightarrow 0$,
respectively, while the solid lines indicate $|a_{\text{AD}}|\rightarrow
\infty$ and $|a_{\text{AT}}|\rightarrow \infty$. All points along
the dashed lines are characterized by reflective 
elastic collisions in the threshold limit. Points along the solid
lines are characterized by reflectionless
collisions and the appearance of a new bound state.  The number of
trimer or tetramer bound states is indicated by $n$ in each
figure. The present calculation smoothly connects the poles and zeroes found along $\lambda=0$ and
$|\lambda|\rightarrow \infty$ in~\cite{Mehta2014PRA}.  

We see from Fig.~\ref{HHLphase} that the atom-dimer
scattering length is typically much larger than the HL scattering length: $a_{\text{AD}}\gg
a$, while the same is not true of $a_\text{AT}$.  We trace the reason
for this back to the behavior of the potential curves in
Fig.~\ref{potCurves}.  First, the trimer (ground-state)
binding is deeper than the HL binding, and therefore the size of the trimer is
correspondingly smaller because it is confined to the deeper portion
of the potential $u(x)$.  This means that the effective atom-trimer
interaction is of shorter range than the atom-dimer interaction, and
typically $a_{\text{AD}}$ is also large in comparison to $a_{\text{AT}}$.

\subsection{Thresholds energies for few-body processes}
\begin{figure}[t]
\begin{center}
\leavevmode
\includegraphics[width=3.0in,clip=true]{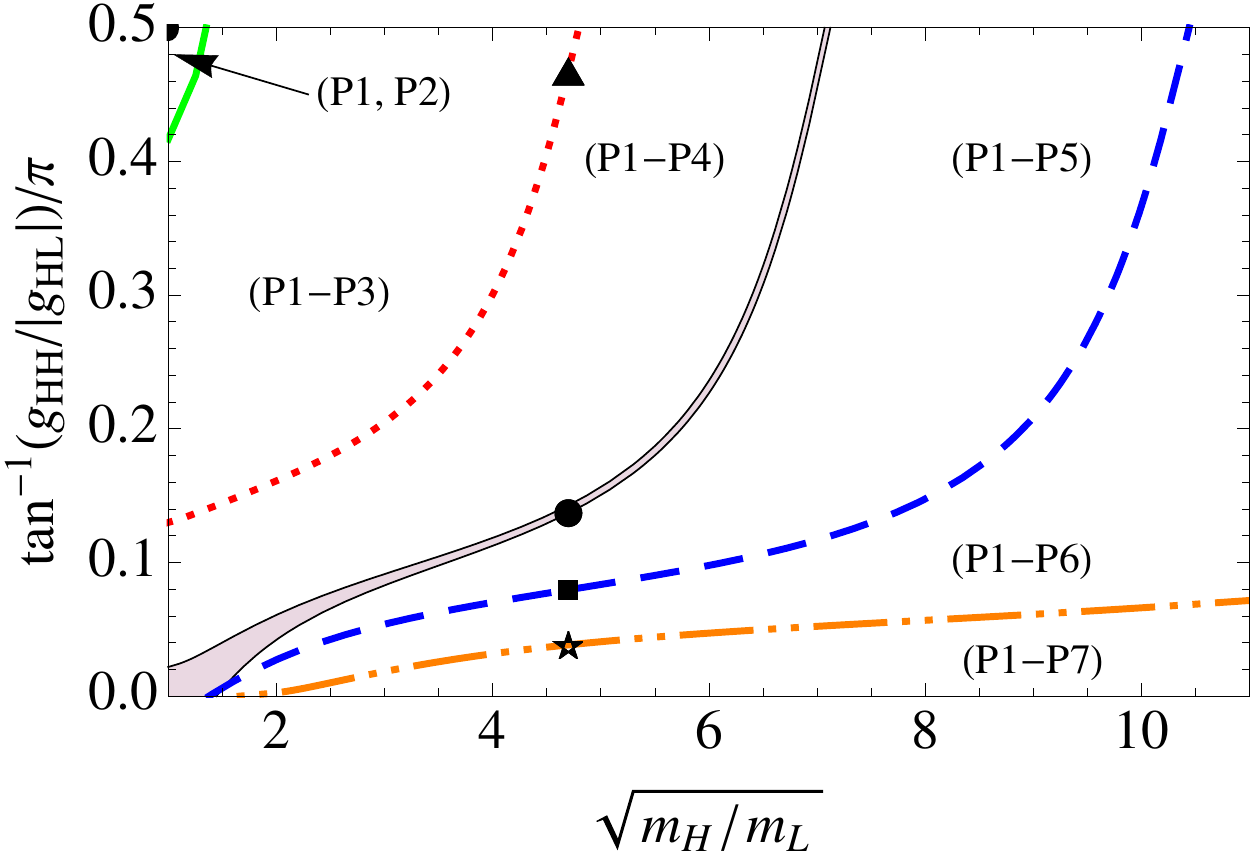}
\caption{(color online) The two-dimensional parameter space is partitioned by lines
  corresponding to the degeneracy of scattering thresholds.  Those
  processes from Table~\ref{Ptable} that are energetically allowed are
  listed in each region.}
\label{pPhaseFinal}
\end{center}
\end{figure}

In Fig.~\ref{pThresholds}, we show the three-body ground
state energy $E_3$ (blue dotted line) and the four-body ground-state energy
$E_4$ (red dashed line), both calculated in the EAA as a function of
$\tan^{-1}{(-\lambda)}/\pi$ for the Li-Cs mass ratio
$\beta^{-1/2}=4.7$.  As one might expect, the energies increase monotonically with
increasing $g_{\text{HH}}$ due to the increasing H-H
repulsion.  We also show the bound state energy $2E_2$ for
two dimers (HL+HL, black line),
$E_2+E_3$ for a dimer plus trimer
(HL+HHL, black dot-dashed line), $E_2+E_4$ for a dimer plus tetramer
(HL+HHHL, purple dot-dot-dashed line), and $2E_3$ for two trimers
(HHL+HHL, orange long-dash line).  Note that for this particular mass ratio,
when $\tan^{-1}{(-\lambda)}< 0.14\pi$, one finds $E_3 <
2E_2$, while for $\tan^{-1}{(-\lambda)} > 0.14\pi$, $E_3 >
2E_2$.  The critical value $\tan^{-1}{(-\lambda)} =
0.14\pi$ (marked by a black circle) indicates the transition point such that
in a gas of dimers, trimer production through the
process $\text{HL}+\text{HL}\rightarrow \text{HHL}+\text{L}$ is
energetically allowed for weaker H-H repulsion.  Placing the critical
value on the plane $\{\beta^{-1/2},\tan^{-1}{(-\lambda)}/\pi\} =
\{4.7,0.14\}$ (see Fig.~\ref{pPhaseFinal}) and repeating the
calculation for other mass ratios allows the parameter space to
be partitioned into regions where this process ((P\ref{pdex}) in
Table~\ref{Ptable}) is either allowed or disallowed. We are able to estimate the error incurred by
the Born-Oppenheimer approximation by numerically calculating the line $E_3 =
2E_2$ both in the EAA and the UAA, drawn as the bottom and top black
lines, respectively, bounding the shaded band in
Fig.~\ref{pPhaseFinal}. We find that the critical mass ratio in the
$|\lambda|\rightarrow  \infty$ limit is $\beta^{-1/2}\approx 7.05$ in the UAA and
$\beta^{-1/2}\approx 7.11$ in the EAA, bracketing the previously quoted
value $\beta^{-1/2}=7.0593$~\cite{Kartavtsev2009JETP}.   

One can imagine other inelastic processes (listed in Table
\ref{Ptable}) involving three, four, five and even six particles that
have thresholds determined purely by the HL, HHL 
and HHHL binding energies. For instance, the rearrangement 
process~(P\ref{ptex}) becomes energetically allowed when
$E_4 \le E_3+E_2$.  For the specific mass ratio $\beta^{-1/2}=4.7$, $E_4=E_3+E_2$
at $\tan{(-\lambda)}=0.08 \pi$ as indicated by the black square on
Fig.~\ref{pThresholds}. For arbitrary values of $\beta^{-1/2}>1$,
process~(P\ref{ptex}) becomes energetically allowed below the dashed blue curve in
Fig.~\ref{pPhaseFinal}.  The three-body recombination processes~(P\ref{p3br}),(P\ref{pd3br}),
and (P\ref{pt3br}) are nearly always energetically allowed for
$\beta^{-1/2}\ge 1$ because there is always an available dimer, trimer
and tetramer state.  The only exception is a small region near the
Tonks-Girardeau limit ($|\lambda|\rightarrow \infty$) where the first
trimer state appears exactly at $\beta^{-1/2}=1$~\cite{Kartavtsev2009JETP} (The
Born-Oppenheimer approximation gives $\beta^{-1/2} \approx
1.08$~\cite{Mehta2014PRA}), and the first tetramer state does 
not appear until $\beta^{-1/2}\gtrsim 1.4$~\cite{Mehta2014PRA}.  Therefore, there is a small region in the
top left corner of Fig.~\ref{pPhaseFinal} (bounded by the solid green line) where process~(P\ref{pt3br})
is not allowed because no tetramer state exists.  The collision of two trimers may produce a tetramer and a dimer
through process~(P\ref{p3342}) only for values of $g_{\text{HH}}$ to
the left of the star in Fig.~\ref{pThresholds} and below the orange
dot-dot-dash line in Fig.~\ref{pPhaseFinal}.  For the range of mass
ratios considered here, (P\ref{p3342}) is not allowed for fermionic H
atoms. Inelastic processes such as these can lead to atom loss and thermal
heating of the trapped gas.  Conversely, atom loss rates can be measured as
a signature of such processes.  

We have labeled each region in Fig.~\ref{pPhaseFinal} by the set of reactions listed in
Table~\ref{Ptable} that are energetically allowed in the forward
direction.  At sufficiently low temperatures, the energy dependence of
the reaction rates for these processes is governed by the corresponding Wigner threshold
law.  For each of the three-body recombination processes
(P\ref{p3br}-P\ref{p3dt}), the threshold law gives
$|S_{\text{fi}}|^2\propto E_{\text{col}}^{\kappa_{\text{min}}}$, where
$\kappa_{\text{min}}$ is the hyperangular quantum number for the
lowest three-body hyperspherical harmonic in the limit of large
hyperradius $R\rightarrow \infty$~\cite{Mehta2007PRA}.  In general,
$\kappa_{\text{min}}$ is an irrational number determined purely by the
masses of the collision partners.  We list the corresponding values of
$\kappa_{\text{min}}$ in Table~\ref{Ptable}.  The energy
dependence of scattering probabilities of two-body processes
(P\ref{pdex}-P\ref{p3342}) is controlled not by the momentum in the
entrance channel, but rather in the fragmentation
channel~\cite{Clark1983PRA,Sadeghpour2000JPB}:
$|S_{\text{fi}}|^2\propto k_f$, and therefore approaches a constant at
threshold if the fragmentation channel lies below the entrance
channel. With the exception of (P\ref{pt3br}), each of
the processes in Table~\ref{Ptable} becomes allowed in the reverse direction on the opposite side of
the corresponding critical line in Fig.~\ref{pPhaseFinal}.

\begin{table}[!t]
\leavevmode
\begin{center}
\begin{tabular}{|l|c|}
\hline
Inelastic process & \shortstack{Threshold law:\\ $|S_\text{fi}|^2\propto E_{\text{col}}^{\kappa_{\text{min}}}$} \\
\hline
\hline

\shortstack{$$\\{(P\refstepcounter{tabitem}\thetabitem)\label{p3br}} $\text{H}+\text{H}+\text{L}
  \rightarrow \text{HL}+\text{H}$\\$$\\$$} 
&\shortstack{$$\\$\kappa_{\text{min}}=\pi/{|\tilde{\phi}-{\pi}/{2}|}$ \\
$\tan{\tilde{\phi}}=\sqrt{\beta/(2+\beta)}$\\$$} \\
\hline

\shortstack{$$\\{(P\refstepcounter{tabitem}\thetabitem)\label{pd3br}} $\text{HL} + \text{H} + \text{H} \rightarrow \text{HHL}+\text{H}$\\$$\\$$}
&\shortstack{$$\\$\kappa_{\text{min}}=\pi/(2|\tilde{\phi}|)$ \\
$\tan{\tilde{\phi}}=\sqrt{{(1+\beta)}/{(3+\beta)}}$\\$$} \\
\hline

\shortstack{$$\\{(P\refstepcounter{tabitem}\thetabitem)\label{pt3br}} $\text{HHL}+\text{H}+\text{H}\rightarrow \text{HHHL}+\text{H}$\\$$\\$$}
&\shortstack{$$\\$\kappa_{\text{min}}={\pi}/{(2|\tilde{\phi}|)}$\\
$\tan{\tilde{\phi}}=\sqrt{(2+\beta)/(4+\beta)}$\\$$} \\
\hline

\shortstack{$$\\{(P\refstepcounter{tabitem}\thetabitem)\label{p3dt}} $\text{HL+HL+H} \rightarrow \text{HHHL +L}$\\$$} &
\shortstack{$$\\$\kappa_{\text{min}}={\pi}/{|\tilde{\phi}-\pi/2|}$\\
$\tan{\tilde{\phi}}=1/\sqrt{3+2\beta}$\\$$}\\ 
\hline

{(P\refstepcounter{tabitem}\thetabitem)\label{pdex}} $\text{HL+HL} \rightarrow \text{HHL + L}$ &
$\kappa_{\text{min}}=0$\\
\hline

{(P\refstepcounter{tabitem}\thetabitem)\label{ptex}} $\text{HHL}+\text{HL} \rightarrow \text{HHHL} + \text{L}$&
$\kappa_{\text{min}}=0$\\
\hline

{(P\refstepcounter{tabitem}\thetabitem)\label{p3342}} $ \text{HHL} +\text{HHL}\rightarrow \text{HHHL}+\text{HL}$&
$\kappa_{\text{min}}=0$\\ 
\hline
\end{tabular}
\end{center}
\caption{Various inelastic processes that are
  energetically allowed in the regions shown in Fig.~\ref{pPhaseFinal}
  are listed here. \label{Ptable} }
\end{table}

\section{Conclusion} 
We have calculated atom-dimer and atom-trimer
scattering lengths, as well as trimer and tetramer energies for the homogeneous 1D
HHL and HHHL systems as a function of the mass ratio
$\beta=m_\text{L}/m_\text{H}$ and coupling ratio
$\lambda=g_{\text{HH}}/g_{\text{HL}}$.  We expect our purely 1D
calculation to be relevant to current experiments with atomic mixtures
in 2D optical lattices in the ``BCS'' limit, when
$a_\perp/a_{\text{3D}}\rightarrow -\infty$ such that the 1D dimer is
weakly bound and has large axial extent compared to $a_\perp$.  When
the few-body system contains only one L impurity, or one ignores the
L-L interaction, all eigenstates are parameterized by the ratios
$\lambda=g_{\text{HH}}/g_{\text{HL}}$ and $\beta={m_L/m_H}$.  We have
determined the regions in this parameter space where certain inelastic
processes involving HL, HHL and HHHL clusters are energetically
allowed, potentially leading to atom loss or heating.  

Finally, we note that the potential curves shown in Fig.~\ref{potCurves} can be used to
calculate the energy-dependent scattering cross section for the
three-body recombination process H+H+HL$\rightarrow$H+HHL.  Such 
calculations are beyond the scope of this paper, but may be pursued in
the future.  

NPM would like to thank D. Blume, J. P. D'Incao and A. G. Volosniev
for insightful discussions.  NPM also thanks the Institute for Nuclear Theory at the
University of Washington for its hospitality and the Department of
Energy for partial support during the completion of this work.

\bibliography{../../AllRefs.bib}

\begin{thebibliography}{37}%
\makeatletter
\providecommand \@ifxundefined [1]{%
 \@ifx{#1\undefined}
}%
\providecommand \@ifnum [1]{%
 \ifnum #1\expandafter \@firstoftwo
 \else \expandafter \@secondoftwo
 \fi
}%
\providecommand \@ifx [1]{%
 \ifx #1\expandafter \@firstoftwo
 \else \expandafter \@secondoftwo
 \fi
}%
\providecommand \natexlab [1]{#1}%
\providecommand \enquote  [1]{``#1''}%
\providecommand \bibnamefont  [1]{#1}%
\providecommand \bibfnamefont [1]{#1}%
\providecommand \citenamefont [1]{#1}%
\providecommand \href@noop [0]{\@secondoftwo}%
\providecommand \href [0]{\begingroup \@sanitize@url \@href}%
\providecommand \@href[1]{\@@startlink{#1}\@@href}%
\providecommand \@@href[1]{\endgroup#1\@@endlink}%
\providecommand \@sanitize@url [0]{\catcode `\\12\catcode `\$12\catcode
  `\&12\catcode `\#12\catcode `\^12\catcode `\_12\catcode `\%12\relax}%
\providecommand \@@startlink[1]{}%
\providecommand \@@endlink[0]{}%
\providecommand \url  [0]{\begingroup\@sanitize@url \@url }%
\providecommand \@url [1]{\endgroup\@href {#1}{\urlprefix }}%
\providecommand \urlprefix  [0]{URL }%
\providecommand \Eprint [0]{\href }%
\providecommand \doibase [0]{http://dx.doi.org/}%
\providecommand \selectlanguage [0]{\@gobble}%
\providecommand \bibinfo  [0]{\@secondoftwo}%
\providecommand \bibfield  [0]{\@secondoftwo}%
\providecommand \translation [1]{[#1]}%
\providecommand \BibitemOpen [0]{}%
\providecommand \bibitemStop [0]{}%
\providecommand \bibitemNoStop [0]{.\EOS\space}%
\providecommand \EOS [0]{\spacefactor3000\relax}%
\providecommand \BibitemShut  [1]{\csname bibitem#1\endcsname}%
\let\auto@bib@innerbib\@empty
\bibitem [{\citenamefont {Cazalilla}\ \emph {et~al.}(2011)\citenamefont
  {Cazalilla}, \citenamefont {Citro}, \citenamefont {Giamarchi}, \citenamefont
  {Orignac},\ and\ \citenamefont {Rigol}}]{Cazalilla2011RMP}%
  \BibitemOpen
  \bibfield  {author} {\bibinfo {author} {\bibfnamefont {M.~A.}\ \bibnamefont
  {Cazalilla}}, \bibinfo {author} {\bibfnamefont {R.}~\bibnamefont {Citro}},
  \bibinfo {author} {\bibfnamefont {T.}~\bibnamefont {Giamarchi}}, \bibinfo
  {author} {\bibfnamefont {E.}~\bibnamefont {Orignac}}, \ and\ \bibinfo
  {author} {\bibfnamefont {M.}~\bibnamefont {Rigol}},\ }\href {\doibase
  10.1103/RevModPhys.83.1405} {\bibfield  {journal} {\bibinfo  {journal} {Rev.
  Mod. Phys.}\ }\textbf {\bibinfo {volume} {83}},\ \bibinfo {pages} {1405}
  (\bibinfo {year} {2011})}\BibitemShut {NoStop}%
\bibitem [{\citenamefont {Guan}\ \emph {et~al.}(2013)\citenamefont {Guan},
  \citenamefont {Batchelor},\ and\ \citenamefont {Lee}}]{Guan2013RMP}%
  \BibitemOpen
  \bibfield  {author} {\bibinfo {author} {\bibfnamefont {X.-W.}\ \bibnamefont
  {Guan}}, \bibinfo {author} {\bibfnamefont {M.~T.}\ \bibnamefont {Batchelor}},
  \ and\ \bibinfo {author} {\bibfnamefont {C.}~\bibnamefont {Lee}},\ }\href
  {\doibase 10.1103/RevModPhys.85.1633} {\bibfield  {journal} {\bibinfo
  {journal} {Rev. Mod. Phys.}\ }\textbf {\bibinfo {volume} {85}},\ \bibinfo
  {pages} {1633} (\bibinfo {year} {2013})}\BibitemShut {NoStop}%
\bibitem [{\citenamefont {Paredes}\ \emph {et~al.}(2004)\citenamefont
  {Paredes}, \citenamefont {Widera}, \citenamefont {Murg}, \citenamefont
  {Mandel}, \citenamefont {F{\"o}lling}, \citenamefont {Cirac}, \citenamefont
  {Shlyapnikov}, \citenamefont {H{\"a}nsch},\ and\ \citenamefont
  {Bloch}}]{Paredes2004Nature}%
  \BibitemOpen
  \bibfield  {author} {\bibinfo {author} {\bibfnamefont {B.}~\bibnamefont
  {Paredes}}, \bibinfo {author} {\bibfnamefont {A.}~\bibnamefont {Widera}},
  \bibinfo {author} {\bibfnamefont {V.}~\bibnamefont {Murg}}, \bibinfo {author}
  {\bibfnamefont {O.}~\bibnamefont {Mandel}}, \bibinfo {author} {\bibfnamefont
  {S.}~\bibnamefont {F{\"o}lling}}, \bibinfo {author} {\bibfnamefont
  {I.}~\bibnamefont {Cirac}}, \bibinfo {author} {\bibfnamefont {G.~V.}\
  \bibnamefont {Shlyapnikov}}, \bibinfo {author} {\bibfnamefont {T.~W.}\
  \bibnamefont {H{\"a}nsch}}, \ and\ \bibinfo {author} {\bibfnamefont
  {I.}~\bibnamefont {Bloch}},\ }\href@noop {} {\bibfield  {journal} {\bibinfo
  {journal} {Nature}\ }\textbf {\bibinfo {volume} {429}},\ \bibinfo {pages}
  {277} (\bibinfo {year} {2004})}\BibitemShut {NoStop}%
\bibitem [{\citenamefont {Moritz}\ \emph {et~al.}(2003)\citenamefont {Moritz},
  \citenamefont {St\"oferle}, \citenamefont {K\"ohl},\ and\ \citenamefont
  {Esslinger}}]{Moritz2003PRL}%
  \BibitemOpen
  \bibfield  {author} {\bibinfo {author} {\bibfnamefont {H.}~\bibnamefont
  {Moritz}}, \bibinfo {author} {\bibfnamefont {T.}~\bibnamefont {St\"oferle}},
  \bibinfo {author} {\bibfnamefont {M.}~\bibnamefont {K\"ohl}}, \ and\ \bibinfo
  {author} {\bibfnamefont {T.}~\bibnamefont {Esslinger}},\ }\href {\doibase
  10.1103/PhysRevLett.91.250402} {\bibfield  {journal} {\bibinfo  {journal}
  {Phys. Rev. Lett.}\ }\textbf {\bibinfo {volume} {91}},\ \bibinfo {pages}
  {250402} (\bibinfo {year} {2003})}\BibitemShut {NoStop}%
\bibitem [{\citenamefont {Kinoshita}\ \emph {et~al.}(2006)\citenamefont
  {Kinoshita}, \citenamefont {Wenger},\ and\ \citenamefont
  {Weiss}}]{Kinoshita2006Nature}%
  \BibitemOpen
  \bibfield  {author} {\bibinfo {author} {\bibfnamefont {T.}~\bibnamefont
  {Kinoshita}}, \bibinfo {author} {\bibfnamefont {T.}~\bibnamefont {Wenger}}, \
  and\ \bibinfo {author} {\bibfnamefont {D.~S.}\ \bibnamefont {Weiss}},\
  }\href@noop {} {\bibfield  {journal} {\bibinfo  {journal} {Nature}\ }\textbf
  {\bibinfo {volume} {440}},\ \bibinfo {pages} {900} (\bibinfo {year}
  {2006})}\BibitemShut {NoStop}%
\bibitem [{\citenamefont {Lieb}\ and\ \citenamefont
  {Liniger}(1963)}]{LiebLiniger1963PhysRev}%
  \BibitemOpen
  \bibfield  {author} {\bibinfo {author} {\bibfnamefont {E.}~\bibnamefont
  {Lieb}}\ and\ \bibinfo {author} {\bibfnamefont {W.}~\bibnamefont {Liniger}},\
  }\href@noop {} {\bibfield  {journal} {\bibinfo  {journal} {Phys. Rev.}\
  }\textbf {\bibinfo {volume} {130}},\ \bibinfo {pages} {1605} (\bibinfo {year}
  {1963})}\BibitemShut {NoStop}%
\bibitem [{\citenamefont {Lieb}(1963)}]{Lieb1963PhysRev}%
  \BibitemOpen
  \bibfield  {author} {\bibinfo {author} {\bibfnamefont {E.~H.}\ \bibnamefont
  {Lieb}},\ }\href {\doibase 10.1103/PhysRev.130.1616} {\bibfield  {journal}
  {\bibinfo  {journal} {Phys. Rev.}\ }\textbf {\bibinfo {volume} {130}},\
  \bibinfo {pages} {1616} (\bibinfo {year} {1963})}\BibitemShut {NoStop}%
\bibitem [{\citenamefont {{J.~B.~McGuire}}(1964)}]{McGuire1964JMP}%
  \BibitemOpen
  \bibfield  {author} {\bibinfo {author} {\bibnamefont {{J.~B.~McGuire}}},\
  }\href@noop {} {\bibfield  {journal} {\bibinfo  {journal} {J. Math. Phys.}\
  }\textbf {\bibinfo {volume} {5}},\ \bibinfo {pages} {622} (\bibinfo {year}
  {1964})}\BibitemShut {NoStop}%
\bibitem [{\citenamefont {Olshanii}(1998)}]{Olshanii1998PRL}%
  \BibitemOpen
  \bibfield  {author} {\bibinfo {author} {\bibfnamefont {M.}~\bibnamefont
  {Olshanii}},\ }\href@noop {} {\bibfield  {journal} {\bibinfo  {journal}
  {Phys. Rev. Lett.}\ }\textbf {\bibinfo {volume} {81}},\ \bibinfo {pages}
  {938} (\bibinfo {year} {1998})}\BibitemShut {NoStop}%
\bibitem [{\citenamefont {Bergeman}\ \emph {et~al.}(2003)\citenamefont
  {Bergeman}, \citenamefont {Moore},\ and\ \citenamefont
  {Olshanii}}]{Bergeman2003PRL}%
  \BibitemOpen
  \bibfield  {author} {\bibinfo {author} {\bibfnamefont {T.}~\bibnamefont
  {Bergeman}}, \bibinfo {author} {\bibfnamefont {M.~G.}\ \bibnamefont {Moore}},
  \ and\ \bibinfo {author} {\bibfnamefont {M.}~\bibnamefont {Olshanii}},\
  }\href@noop {} {\bibfield  {journal} {\bibinfo  {journal} {Phys. Rev. Lett.}\
  }\textbf {\bibinfo {volume} {91}},\ \bibinfo {pages} {163201} (\bibinfo
  {year} {2003})}\BibitemShut {NoStop}%
\bibitem [{\citenamefont {Gradshteyn}\ and\ \citenamefont
  {Ryzhik}(2014)}]{GradshteynRyzhik}%
  \BibitemOpen
  \bibfield  {author} {\bibinfo {author} {\bibfnamefont {I.~S.}\ \bibnamefont
  {Gradshteyn}}\ and\ \bibinfo {author} {\bibfnamefont {I.~M.}\ \bibnamefont
  {Ryzhik}},\ }\href@noop {} {\emph {\bibinfo {title} {Table of integrals,
  series and products}}},\ \bibinfo {edition} {7th}\ ed.,\ edited by\ \bibinfo
  {editor} {\bibfnamefont {A.}~\bibnamefont {Jeffrey}}\ and\ \bibinfo {editor}
  {\bibfnamefont {D.}~\bibnamefont {Zwillinger}}\ (\bibinfo  {publisher} {New
  York: Academic Press},\ \bibinfo {year} {2014})\BibitemShut {NoStop}%
\bibitem [{\citenamefont {Astrakharchik}\ \emph
  {et~al.}(2004{\natexlab{a}})\citenamefont {Astrakharchik}, \citenamefont
  {Blume}, \citenamefont {Giorgini},\ and\ \citenamefont
  {Granger}}]{Astrakharchik2004PRL}%
  \BibitemOpen
  \bibfield  {author} {\bibinfo {author} {\bibfnamefont {G.~E.}\ \bibnamefont
  {Astrakharchik}}, \bibinfo {author} {\bibfnamefont {D.}~\bibnamefont
  {Blume}}, \bibinfo {author} {\bibfnamefont {S.}~\bibnamefont {Giorgini}}, \
  and\ \bibinfo {author} {\bibfnamefont {B.~E.}\ \bibnamefont {Granger}},\
  }\href {\doibase 10.1103/PhysRevLett.92.030402} {\bibfield  {journal}
  {\bibinfo  {journal} {Phys. Rev. Lett.}\ }\textbf {\bibinfo {volume} {92}},\
  \bibinfo {pages} {030402} (\bibinfo {year} {2004}{\natexlab{a}})}\BibitemShut
  {NoStop}%
\bibitem [{\citenamefont {Astrakharchik}\ \emph
  {et~al.}(2004{\natexlab{b}})\citenamefont {Astrakharchik}, \citenamefont
  {Blume}, \citenamefont {Giorgini},\ and\ \citenamefont
  {Granger}}]{Astrakharchik2004JPhysB}%
  \BibitemOpen
  \bibfield  {author} {\bibinfo {author} {\bibfnamefont {G.~E.}\ \bibnamefont
  {Astrakharchik}}, \bibinfo {author} {\bibfnamefont {D.}~\bibnamefont
  {Blume}}, \bibinfo {author} {\bibfnamefont {S.}~\bibnamefont {Giorgini}}, \
  and\ \bibinfo {author} {\bibfnamefont {B.~E.}\ \bibnamefont {Granger}},\
  }\href {http://stacks.iop.org/0953-4075/37/i=7/a=066} {\bibfield  {journal}
  {\bibinfo  {journal} {Journal of Physics B: Atomic, Molecular and Optical
  Physics}\ }\textbf {\bibinfo {volume} {37}},\ \bibinfo {pages} {S205}
  (\bibinfo {year} {2004}{\natexlab{b}})}\BibitemShut {NoStop}%
\bibitem [{\citenamefont {Moritz}\ \emph {et~al.}(2005)\citenamefont {Moritz},
  \citenamefont {St\"oferle}, \citenamefont {G\"unter}, \citenamefont
  {K\"ohl},\ and\ \citenamefont {Esslinger}}]{Moritz2005PRL}%
  \BibitemOpen
  \bibfield  {author} {\bibinfo {author} {\bibfnamefont {H.}~\bibnamefont
  {Moritz}}, \bibinfo {author} {\bibfnamefont {T.}~\bibnamefont {St\"oferle}},
  \bibinfo {author} {\bibfnamefont {K.}~\bibnamefont {G\"unter}}, \bibinfo
  {author} {\bibfnamefont {M.}~\bibnamefont {K\"ohl}}, \ and\ \bibinfo {author}
  {\bibfnamefont {T.}~\bibnamefont {Esslinger}},\ }\href {\doibase
  10.1103/PhysRevLett.94.210401} {\bibfield  {journal} {\bibinfo  {journal}
  {Phys. Rev. Lett.}\ }\textbf {\bibinfo {volume} {94}},\ \bibinfo {pages}
  {210401} (\bibinfo {year} {2005})}\BibitemShut {NoStop}%
\bibitem [{\citenamefont {Mora}\ \emph
  {et~al.}(2005{\natexlab{a}})\citenamefont {Mora}, \citenamefont {Egger},\
  and\ \citenamefont {Gogolin}}]{Mora2005PRA}%
  \BibitemOpen
  \bibfield  {author} {\bibinfo {author} {\bibfnamefont {C.}~\bibnamefont
  {Mora}}, \bibinfo {author} {\bibfnamefont {R.}~\bibnamefont {Egger}}, \ and\
  \bibinfo {author} {\bibfnamefont {A.~O.}\ \bibnamefont {Gogolin}},\
  }\href@noop {} {\bibfield  {journal} {\bibinfo  {journal} {Phys. Rev. A}\
  }\textbf {\bibinfo {volume} {71}},\ \bibinfo {pages} {052705} (\bibinfo
  {year} {2005}{\natexlab{a}})}\BibitemShut {NoStop}%
\bibitem [{\citenamefont {Mora}\ \emph
  {et~al.}(2005{\natexlab{b}})\citenamefont {Mora}, \citenamefont {Komnik},
  \citenamefont {Egger},\ and\ \citenamefont {Gogolin}}]{Mora2005PRL}%
  \BibitemOpen
  \bibfield  {author} {\bibinfo {author} {\bibfnamefont {C.}~\bibnamefont
  {Mora}}, \bibinfo {author} {\bibfnamefont {A.}~\bibnamefont {Komnik}},
  \bibinfo {author} {\bibfnamefont {R.}~\bibnamefont {Egger}}, \ and\ \bibinfo
  {author} {\bibfnamefont {A.~O.}\ \bibnamefont {Gogolin}},\ }\href {\doibase
  10.1103/PhysRevLett.95.080403} {\bibfield  {journal} {\bibinfo  {journal}
  {Phys. Rev. Lett.}\ }\textbf {\bibinfo {volume} {95}},\ \bibinfo {pages}
  {080403} (\bibinfo {year} {2005}{\natexlab{b}})}\BibitemShut {NoStop}%
\bibitem [{\citenamefont {Wenz}\ \emph {et~al.}(2013)\citenamefont {Wenz},
  \citenamefont {Z{\"u}rn}, \citenamefont {Murmann}, \citenamefont {Brouzos},
  \citenamefont {Lompe},\ and\ \citenamefont {Jochim}}]{Wenz2013Science}%
  \BibitemOpen
  \bibfield  {author} {\bibinfo {author} {\bibfnamefont {A.}~\bibnamefont
  {Wenz}}, \bibinfo {author} {\bibfnamefont {G.}~\bibnamefont {Z{\"u}rn}},
  \bibinfo {author} {\bibfnamefont {S.}~\bibnamefont {Murmann}}, \bibinfo
  {author} {\bibfnamefont {I.}~\bibnamefont {Brouzos}}, \bibinfo {author}
  {\bibfnamefont {T.}~\bibnamefont {Lompe}}, \ and\ \bibinfo {author}
  {\bibfnamefont {S.}~\bibnamefont {Jochim}},\ }\href@noop {} {\bibfield
  {journal} {\bibinfo  {journal} {Science}\ }\textbf {\bibinfo {volume}
  {342}},\ \bibinfo {pages} {457} (\bibinfo {year} {2013})}\BibitemShut
  {NoStop}%
\bibitem [{\citenamefont {Catani}\ \emph {et~al.}(2012)\citenamefont {Catani},
  \citenamefont {Lamporesi}, \citenamefont {Naik}, \citenamefont {Gring},
  \citenamefont {Inguscio}, \citenamefont {Minardi}, \citenamefont {Kantian},\
  and\ \citenamefont {Giamarchi}}]{Catani2012PRA}%
  \BibitemOpen
  \bibfield  {author} {\bibinfo {author} {\bibfnamefont {J.}~\bibnamefont
  {Catani}}, \bibinfo {author} {\bibfnamefont {G.}~\bibnamefont {Lamporesi}},
  \bibinfo {author} {\bibfnamefont {D.}~\bibnamefont {Naik}}, \bibinfo {author}
  {\bibfnamefont {M.}~\bibnamefont {Gring}}, \bibinfo {author} {\bibfnamefont
  {M.}~\bibnamefont {Inguscio}}, \bibinfo {author} {\bibfnamefont
  {F.}~\bibnamefont {Minardi}}, \bibinfo {author} {\bibfnamefont
  {A.}~\bibnamefont {Kantian}}, \ and\ \bibinfo {author} {\bibfnamefont
  {T.}~\bibnamefont {Giamarchi}},\ }\href@noop {} {\bibfield  {journal}
  {\bibinfo  {journal} {Phys. Rev. A}\ }\textbf {\bibinfo {volume} {85}},\
  \bibinfo {pages} {023623} (\bibinfo {year} {2012})}\BibitemShut {NoStop}%
\bibitem [{\citenamefont {Pagano}\ \emph {et~al.}(2014)\citenamefont {Pagano},
  \citenamefont {Mancini}, \citenamefont {Cappellini}, \citenamefont
  {Lombardi}, \citenamefont {Sch{\"a}fer}, \citenamefont {Hu}, \citenamefont
  {Liu}, \citenamefont {Catani}, \citenamefont {Sias}, \citenamefont
  {Inguscio}, \citenamefont {Massimo},\ and\ \citenamefont
  {Fallani}}]{Pagano2014NatPhys}%
  \BibitemOpen
  \bibfield  {author} {\bibinfo {author} {\bibfnamefont {G.}~\bibnamefont
  {Pagano}}, \bibinfo {author} {\bibfnamefont {M.}~\bibnamefont {Mancini}},
  \bibinfo {author} {\bibfnamefont {G.}~\bibnamefont {Cappellini}}, \bibinfo
  {author} {\bibfnamefont {P.}~\bibnamefont {Lombardi}}, \bibinfo {author}
  {\bibfnamefont {F.}~\bibnamefont {Sch{\"a}fer}}, \bibinfo {author}
  {\bibfnamefont {H.}~\bibnamefont {Hu}}, \bibinfo {author} {\bibfnamefont
  {X.-J.}\ \bibnamefont {Liu}}, \bibinfo {author} {\bibfnamefont
  {J.}~\bibnamefont {Catani}}, \bibinfo {author} {\bibfnamefont
  {C.}~\bibnamefont {Sias}}, \bibinfo {author} {\bibfnamefont {M.}~\bibnamefont
  {Inguscio}}, \bibinfo {author} {\bibfnamefont {I.}~\bibnamefont {Massimo}}, \
  and\ \bibinfo {author} {\bibfnamefont {L.}~\bibnamefont {Fallani}},\
  }\href@noop {} {\bibfield  {journal} {\bibinfo  {journal} {Nature Physics}\
  }\textbf {\bibinfo {volume} {10}},\ \bibinfo {pages} {198} (\bibinfo {year}
  {2014})}\BibitemShut {NoStop}%
\bibitem [{\citenamefont {Gharashi}\ and\ \citenamefont
  {Blume}(2013)}]{Gharashi2013PRL}%
  \BibitemOpen
  \bibfield  {author} {\bibinfo {author} {\bibfnamefont {S.~E.}\ \bibnamefont
  {Gharashi}}\ and\ \bibinfo {author} {\bibfnamefont {D.}~\bibnamefont
  {Blume}},\ }\href {\doibase 10.1103/PhysRevLett.111.045302} {\bibfield
  {journal} {\bibinfo  {journal} {Phys. Rev. Lett.}\ }\textbf {\bibinfo
  {volume} {111}},\ \bibinfo {pages} {045302} (\bibinfo {year}
  {2013})}\BibitemShut {NoStop}%
\bibitem [{\citenamefont {Gharashi}\ \emph {et~al.}(2012)\citenamefont
  {Gharashi}, \citenamefont {Daily},\ and\ \citenamefont
  {Blume}}]{Gharashi2012PRA}%
  \BibitemOpen
  \bibfield  {author} {\bibinfo {author} {\bibfnamefont {S.~E.}\ \bibnamefont
  {Gharashi}}, \bibinfo {author} {\bibfnamefont {K.~M.}\ \bibnamefont {Daily}},
  \ and\ \bibinfo {author} {\bibfnamefont {D.}~\bibnamefont {Blume}},\ }\href
  {\doibase 10.1103/PhysRevA.86.042702} {\bibfield  {journal} {\bibinfo
  {journal} {Phys. Rev. A}\ }\textbf {\bibinfo {volume} {86}},\ \bibinfo
  {pages} {042702} (\bibinfo {year} {2012})}\BibitemShut {NoStop}%
\bibitem [{\citenamefont {Sowi\ifmmode~\acute{n}\else \'{n}\fi{}ski}\ \emph
  {et~al.}(2013)\citenamefont {Sowi\ifmmode~\acute{n}\else \'{n}\fi{}ski},
  \citenamefont {Grass}, \citenamefont {Dutta},\ and\ \citenamefont
  {Lewenstein}}]{Sowinski2013PRA}%
  \BibitemOpen
  \bibfield  {author} {\bibinfo {author} {\bibfnamefont {T.}~\bibnamefont
  {Sowi\ifmmode~\acute{n}\else \'{n}\fi{}ski}}, \bibinfo {author}
  {\bibfnamefont {T.}~\bibnamefont {Grass}}, \bibinfo {author} {\bibfnamefont
  {O.}~\bibnamefont {Dutta}}, \ and\ \bibinfo {author} {\bibfnamefont
  {M.}~\bibnamefont {Lewenstein}},\ }\href {\doibase
  10.1103/PhysRevA.88.033607} {\bibfield  {journal} {\bibinfo  {journal} {Phys.
  Rev. A}\ }\textbf {\bibinfo {volume} {88}},\ \bibinfo {pages} {033607}
  (\bibinfo {year} {2013})}\BibitemShut {NoStop}%
\bibitem [{\citenamefont {Gharashi}\ \emph {et~al.}(2015)\citenamefont
  {Gharashi}, \citenamefont {Yin}, \citenamefont {Yan},\ and\ \citenamefont
  {Blume}}]{Gharashi2015PRA}%
  \BibitemOpen
  \bibfield  {author} {\bibinfo {author} {\bibfnamefont {S.~E.}\ \bibnamefont
  {Gharashi}}, \bibinfo {author} {\bibfnamefont {X.~Y.}\ \bibnamefont {Yin}},
  \bibinfo {author} {\bibfnamefont {Y.}~\bibnamefont {Yan}}, \ and\ \bibinfo
  {author} {\bibfnamefont {D.}~\bibnamefont {Blume}},\ }\href {\doibase
  10.1103/PhysRevA.91.013620} {\bibfield  {journal} {\bibinfo  {journal} {Phys.
  Rev. A}\ }\textbf {\bibinfo {volume} {91}},\ \bibinfo {pages} {013620}
  (\bibinfo {year} {2015})}\BibitemShut {NoStop}%
\bibitem [{\citenamefont {Lindgren}\ \emph {et~al.}(2014)\citenamefont
  {Lindgren}, \citenamefont {Rotureau}, \citenamefont {Forss{\'e}n},
  \citenamefont {Volosniev},\ and\ \citenamefont {Zinner}}]{Lindgren2014NJP}%
  \BibitemOpen
  \bibfield  {author} {\bibinfo {author} {\bibfnamefont {E.}~\bibnamefont
  {Lindgren}}, \bibinfo {author} {\bibfnamefont {J.}~\bibnamefont {Rotureau}},
  \bibinfo {author} {\bibfnamefont {C.}~\bibnamefont {Forss{\'e}n}}, \bibinfo
  {author} {\bibfnamefont {A.}~\bibnamefont {Volosniev}}, \ and\ \bibinfo
  {author} {\bibfnamefont {N.~T.}\ \bibnamefont {Zinner}},\ }\href@noop {}
  {\bibfield  {journal} {\bibinfo  {journal} {New Journal of Physics}\ }\textbf
  {\bibinfo {volume} {16}},\ \bibinfo {pages} {063003} (\bibinfo {year}
  {2014})}\BibitemShut {NoStop}%
\bibitem [{\citenamefont {Volosniev}\ \emph {et~al.}(2014)\citenamefont
  {Volosniev}, \citenamefont {Fedorov}, \citenamefont {Jensen}, \citenamefont
  {Valiente},\ and\ \citenamefont {Zinner}}]{Volosniev2014NatureComm}%
  \BibitemOpen
  \bibfield  {author} {\bibinfo {author} {\bibfnamefont {A.}~\bibnamefont
  {Volosniev}}, \bibinfo {author} {\bibfnamefont {D.~V.}\ \bibnamefont
  {Fedorov}}, \bibinfo {author} {\bibfnamefont {A.~S.}\ \bibnamefont {Jensen}},
  \bibinfo {author} {\bibfnamefont {M.}~\bibnamefont {Valiente}}, \ and\
  \bibinfo {author} {\bibfnamefont {N.~T.}\ \bibnamefont {Zinner}},\
  }\href@noop {} {\bibfield  {journal} {\bibinfo  {journal} {Nature
  communications}\ }\textbf {\bibinfo {volume} {5}},\ \bibinfo {pages} {1}
  (\bibinfo {year} {2014})}\BibitemShut {NoStop}%
\bibitem [{\citenamefont {Levinsen}\ \emph {et~al.}(2015)\citenamefont
  {Levinsen}, \citenamefont {Massignan}, \citenamefont {Bruun},\ and\
  \citenamefont {Parish}}]{Levinsen2015ScienceAdvances}%
  \BibitemOpen
  \bibfield  {author} {\bibinfo {author} {\bibfnamefont {J.}~\bibnamefont
  {Levinsen}}, \bibinfo {author} {\bibfnamefont {P.}~\bibnamefont {Massignan}},
  \bibinfo {author} {\bibfnamefont {G.~M.}\ \bibnamefont {Bruun}}, \ and\
  \bibinfo {author} {\bibfnamefont {M.~M.}\ \bibnamefont {Parish}},\
  }\href@noop {} {\bibfield  {journal} {\bibinfo  {journal} {Science Advances}\
  }\textbf {\bibinfo {volume} {1}},\ \bibinfo {pages} {e1500197} (\bibinfo
  {year} {2015})}\BibitemShut {NoStop}%
\bibitem [{\citenamefont {Dehkharghani}\ \emph {et~al.}(2015)\citenamefont
  {Dehkharghani}, \citenamefont {Volosniev},\ and\ \citenamefont
  {Zinner}}]{Dehkharghani2015Preprint}%
  \BibitemOpen
  \bibfield  {author} {\bibinfo {author} {\bibfnamefont {A.~S.}\ \bibnamefont
  {Dehkharghani}}, \bibinfo {author} {\bibfnamefont {A.~G.}\ \bibnamefont
  {Volosniev}}, \ and\ \bibinfo {author} {\bibfnamefont {N.~T.}\ \bibnamefont
  {Zinner}},\ }\href@noop {} {\  (\bibinfo {year} {2015})},\ \Eprint
  {http://arxiv.org/abs/1503.03725} {arXiv:1503.03725 [cond-mat.quant-gas]}
  \BibitemShut {NoStop}%
\bibitem [{\citenamefont {Girardeau}(1960)}]{Girardeau1960JMP}%
  \BibitemOpen
  \bibfield  {author} {\bibinfo {author} {\bibfnamefont {M.~D.}\ \bibnamefont
  {Girardeau}},\ }\href@noop {} {\bibfield  {journal} {\bibinfo  {journal}
  {J.~Math Phys.}\ }\textbf {\bibinfo {volume} {1}},\ \bibinfo {pages} {516}
  (\bibinfo {year} {1960})}\BibitemShut {NoStop}%
\bibitem [{\citenamefont {Kartavtsev}\ \emph {et~al.}(2009)\citenamefont
  {Kartavtsev}, \citenamefont {Malykh},\ and\ \citenamefont
  {Sofianos}}]{Kartavtsev2009JETP}%
  \BibitemOpen
  \bibfield  {author} {\bibinfo {author} {\bibfnamefont {O.~I.}\ \bibnamefont
  {Kartavtsev}}, \bibinfo {author} {\bibfnamefont {A.~V.}\ \bibnamefont
  {Malykh}}, \ and\ \bibinfo {author} {\bibfnamefont {S.~A.}\ \bibnamefont
  {Sofianos}},\ }\href@noop {} {\bibfield  {journal} {\bibinfo  {journal}
  {ZhETF}\ }\textbf {\bibinfo {volume} {135}},\ \bibinfo {pages} {419}
  (\bibinfo {year} {2009})}\BibitemShut {NoStop}%
\bibitem [{\citenamefont {Mehta}(2014)}]{Mehta2014PRA}%
  \BibitemOpen
  \bibfield  {author} {\bibinfo {author} {\bibfnamefont {N.~P.}\ \bibnamefont
  {Mehta}},\ }\href {\doibase 10.1103/PhysRevA.89.052706} {\bibfield  {journal}
  {\bibinfo  {journal} {Phys. Rev. A}\ }\textbf {\bibinfo {volume} {89}},\
  \bibinfo {pages} {052706} (\bibinfo {year} {2014})}\BibitemShut {NoStop}%
\bibitem [{\citenamefont {Brattsev}(1965)}]{Brattsev1965Dokl}%
  \BibitemOpen
  \bibfield  {author} {\bibinfo {author} {\bibfnamefont {V.~F.}\ \bibnamefont
  {Brattsev}},\ }\href@noop {} {\bibfield  {journal} {\bibinfo  {journal}
  {Sov.~Phys.~Dokl.}\ }\textbf {\bibinfo {volume} {10}},\ \bibinfo {pages} {44}
  (\bibinfo {year} {1965})}\BibitemShut {NoStop}%
\bibitem [{\citenamefont {Epstein}(1966)}]{Epstein1966JCP}%
  \BibitemOpen
  \bibfield  {author} {\bibinfo {author} {\bibfnamefont {S.~T.}\ \bibnamefont
  {Epstein}},\ }\href@noop {} {\bibfield  {journal} {\bibinfo  {journal}
  {J.~Chem.~Phys.}\ }\textbf {\bibinfo {volume} {44}},\ \bibinfo {pages} {836}
  (\bibinfo {year} {1966})}\BibitemShut {NoStop}%
\bibitem [{\citenamefont {Starace}\ and\ \citenamefont
  {Webster}(1979)}]{Strace1979PRA}%
  \BibitemOpen
  \bibfield  {author} {\bibinfo {author} {\bibfnamefont {A.~F.}\ \bibnamefont
  {Starace}}\ and\ \bibinfo {author} {\bibfnamefont {G.~L.}\ \bibnamefont
  {Webster}},\ }\href {\doibase 10.1103/PhysRevA.19.1629} {\bibfield  {journal}
  {\bibinfo  {journal} {Phys. Rev. A}\ }\textbf {\bibinfo {volume} {19}},\
  \bibinfo {pages} {1629} (\bibinfo {year} {1979})}\BibitemShut {NoStop}%
\bibitem [{\citenamefont {Coelho}\ and\ \citenamefont
  {Hornos}(1991)}]{Coelho1991PRA}%
  \BibitemOpen
  \bibfield  {author} {\bibinfo {author} {\bibfnamefont {H.~T.}\ \bibnamefont
  {Coelho}}\ and\ \bibinfo {author} {\bibfnamefont {J.~E.}\ \bibnamefont
  {Hornos}},\ }\href {\doibase 10.1103/PhysRevA.43.6379} {\bibfield  {journal}
  {\bibinfo  {journal} {Phys. Rev. A}\ }\textbf {\bibinfo {volume} {43}},\
  \bibinfo {pages} {6379} (\bibinfo {year} {1991})}\BibitemShut {NoStop}%
\bibitem [{\citenamefont {Mehta}\ \emph {et~al.}(2007)\citenamefont {Mehta},
  \citenamefont {Esry},\ and\ \citenamefont {Greene}}]{Mehta2007PRA}%
  \BibitemOpen
  \bibfield  {author} {\bibinfo {author} {\bibfnamefont {N.~P.}\ \bibnamefont
  {Mehta}}, \bibinfo {author} {\bibfnamefont {B.~D.}\ \bibnamefont {Esry}}, \
  and\ \bibinfo {author} {\bibfnamefont {C.~H.}\ \bibnamefont {Greene}},\
  }\href {\doibase 10.1103/PhysRevA.76.022711} {\bibfield  {journal} {\bibinfo
  {journal} {Phys. Rev. A}\ }\textbf {\bibinfo {volume} {76}},\ \bibinfo
  {pages} {022711} (\bibinfo {year} {2007})}\BibitemShut {NoStop}%
\bibitem [{\citenamefont {Clark}(1983)}]{Clark1983PRA}%
  \BibitemOpen
  \bibfield  {author} {\bibinfo {author} {\bibfnamefont {C.~W.}\ \bibnamefont
  {Clark}},\ }\href {\doibase 10.1103/PhysRevA.28.83} {\bibfield  {journal}
  {\bibinfo  {journal} {Phys. Rev. A}\ }\textbf {\bibinfo {volume} {28}},\
  \bibinfo {pages} {83} (\bibinfo {year} {1983})}\BibitemShut {NoStop}%
\bibitem [{\citenamefont {Sadeghpour}\ \emph {et~al.}(2000)\citenamefont
  {Sadeghpour}, \citenamefont {Bohn}, \citenamefont {Cavagnero}, \citenamefont
  {Esry}, \citenamefont {Fabrikant}, \citenamefont {Macek},\ and\ \citenamefont
  {Rau}}]{Sadeghpour2000JPB}%
  \BibitemOpen
  \bibfield  {author} {\bibinfo {author} {\bibfnamefont {H.~R.}\ \bibnamefont
  {Sadeghpour}}, \bibinfo {author} {\bibfnamefont {J.~L.}\ \bibnamefont
  {Bohn}}, \bibinfo {author} {\bibfnamefont {M.~J.}\ \bibnamefont {Cavagnero}},
  \bibinfo {author} {\bibfnamefont {B.~D.}\ \bibnamefont {Esry}}, \bibinfo
  {author} {\bibfnamefont {I.~I.}\ \bibnamefont {Fabrikant}}, \bibinfo {author}
  {\bibfnamefont {J.~H.}\ \bibnamefont {Macek}}, \ and\ \bibinfo {author}
  {\bibfnamefont {A.~R.~P.}\ \bibnamefont {Rau}},\ }\href
  {http://stacks.iop.org/0953-4075/33/i=5/a=201} {\bibfield  {journal}
  {\bibinfo  {journal} {Journal of Physics B: Atomic, Molecular and Optical
  Physics}\ }\textbf {\bibinfo {volume} {33}},\ \bibinfo {pages} {R93}
  (\bibinfo {year} {2000})}\BibitemShut {NoStop}%
\end{thebibliography}%

\end{document}